# Carbon Capture from wet vapors

Silvina Gatica[1]

[1]Department of Physics and Astronomy, Howard University, 2355 Sixth St NW,

Washington, DC 20059, USA. December 15, 2025

## Abstract

We present molecular dynamics simulations of the adsorption of mixed $CO_2$–water vapors on a graphene-flakes (GF) substrate, a model inspired by the microporous structure of activated carbons. Adsorption strength is quantified through a reduced energy measure that avoids ambiguities associated with defining adsorption regions.

We find that $CO_2$ adsorbs more strongly and more rapidly than water across all temperatures studied. Adsorption from wet vapors was observed to result in higher $CO_2$ uptake compared with dry vapors at temperatures below 375 K. At intermediate temperatures, water molecules were seen to form clusters that interact with both the substrate and $CO_2$, potentially promoting $CO_2$ adsorption.

Additionally, we observe that the GF substrate inhibits the formation of large water clusters, altering water aggregation dynamics near the surface. These findings highlight a cooperative adsorption mecha- nism in humid environments and suggest that graphene-flakes–based materials may perform effectively for carbon capture under realistic, moisture-containing conditions.

## 1. Introduction

Carbon Dioxide ($CO_2$) is one of the most prevalent and concerning greenhouse gasses that contribute to global warming, driven primarily by human activities such as industrial processes, the burning of fossil fuels, deforestation, and agriculture [1–3]. As the concentration of $CO_2$ in the atmosphere continues to increase, it exacerbates the greenhouse effect, leading to climate change, increased global temperatures, and a cascade of other environmental challenges. Reducing the emission of $CO_2$ into the atmosphere is therefore of critical importance in mitigating the harmful effects of climate change. Efforts to address this issue generally focus on two major strategies: reducing emissions at the source and enhancing the sequestration of $CO_2$ through various environmental and technological approaches [4].

One of the most effective ways to mitigate the environmental impact of $CO_2$ is to separate $CO_2$ from other gasses such as methane ($CH_4$) and nitrogen ($N_2$), particularly in industrial settings such as power plants, where these gasses are often present in flue gasses. Separation of $CO_2$ is a critical step in enabling both the capture and storage of carbon, which can help prevent its release into the atmosphere. In this context, selective-adsorption emerges as a promising method for gas separation, offering a reliable and efficient technique to isolate $CO_2$ from other gasses in a mixture. This technique utilizes the differential interactions between the gas molecules and the surface of the adsorbent material to selectively capture one gas over others. [5]

Given the importance of $CO_2$ separation, there has been considerable research into the selective adsorption of $CO_2$, $CH_4$, and $N_2$ mixtures using various adsorbent materials. Among the most widely studied materials are Metal-Organic Frameworks (MOFs) and Activated Carbons (ACs), which are highly porous and have high surface areas that make them ideal candidates for gas adsorption [6–15].

Activated carbons offer several advantages for carbon capture, primarily due to their high surface area and excellent adsorption capacity. With surface areas ranging from 1000 to 3000 $m^2/g$, ACs provide numerous sites for carbon dioxide molecules to adsorb, making them highly effective for gas capture applications. Their ability to adsorb large quantities of $CO_2$ is essential for industrial carbon capture, where large volumes of gas need to be processed. Additionally, activated carbons are highly versatile and can be tailored for specific applications. By adjusting the activation process or incorporating other materials, their adsorption properties can be optimized for different operating conditions, such as varying temperatures and pressures [16–18].

Another key advantage of activated carbons is their reusability, which is critical for long-term cost-effectiveness. ACs can be regenerated through physical or chemical methods, allowing them to be reused for multiple cycles of $CO_2$ capture. Moreover, activated carbons are relatively inexpensive, especially when derived from renewable biomass sources like coconut shells or wood, making them an attractive option for large-scale carbon capture operations. Their use in carbon capture is also supported by a mature technology base, as activated carbons have been employed for decades in air filtration and gas separation. [13,14].

We recently proposed a substrate designed with graphene flakes (GF) as an adsorbent material inspired by the characteristics of ACs derived from wood sawdust and tropical fruit seeds, as outlined by Gomez-Delgado et al. [13]. In our study, we computed the adsorption isotherms for $N_2$ at 77 K and $CO_2$ at 273 K on the GF substrate, and the results showed similarities to those of ACs, indicating that our model can capture key aspects of AC-based substrates. [5]

Humidity can influence carbon capture in both natural and industrial processes. In industrial carbon capture technologies, humidity can also affect the performance of $CO_2$ capture materials, such as amine- based solvents and chemical scrubbers. High humidity can interact with solvents, reducing their ability to effectively absorb carbon dioxide [19]. Although some emerging technologies, such as MOFs, are being developed to withstand higher humidity levels, the general impact of humidity on carbon capture remains complex. High humidity does not directly increase carbon capture in these systems and can instead reduce their efficiency under certain conditions [20,21].

Magnin et al. investigated the effect of humidity on carbon capture in MOFs. [22] They found that in some regime the electric field of water acts as a supplemental driving force, while in other cases it becomes a competitor, decreasing the carbon capture capacity.

In this work, we report simulations of $CO_2$ and water mixed-vapor adsorption on the GF substrate, focusing on the influence of water on $CO_2$ uptake.

This article is organized as follows. Section 2 describes the model and simulation method. In Section 3 we discuss the results, and in Section 4 we present our summary and conclusions.

**2. Model and Method**

We run MD simulations in the NVT ensemble using the Nose Hoover thermostat. The timestep is 0.5 ps, and the simulations are run using the LAMMPS platform. [23]

The simulation cell is rectangular with dimensions 20 x 20 × 22.8 nm with periodic boundary conditions in all directions. The initial configuration consists of 120 water molecules plus 120 $CO_2$ molecules located on a square lattice. The vapor is equilibrated at the desired temperature during one million steps, after which the cell is expanded, and the molecules are rigidly shifted to make room for the substrate that is inserted in the middle of the cell. (see Fig. 1) The substrate is rigid and it is kept frozen at all times, and it is excluded from the temperature calculation, i.e. it only acts as an interacting force.

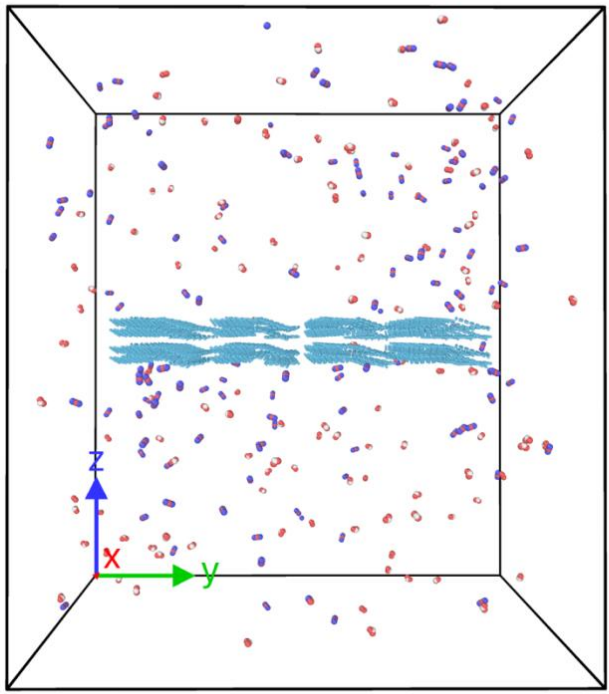

Figure 1: Simulation cell with the substrate (cyan). The atoms are not in scale.

The $CO_2$ molecule is modeled as a linear rigid body with three LJ sites and three partial charges on each atom. The bond length is $b = 0.116$ nm; the carbon atom has a positive charge $q_C = +0.7e$, and the oxygen atoms have a negative charge $q_O = -0.35e$ [24].

The water molecule is modeled as a rigid body with OH bond 0.1nm and HOH angle 109.47°. We adopt the SPC/E force [25] parameters as listed in Table 1.

The intermolecular interaction energy is computed as a combination of the LJ and Coulomb potentials between partial charges.

We adopt the LJ parameters from Ref. [24] (see Table 1). The LJ parameters of the carbon atoms in the substrate are $\sigma = 0.34$ nm, $\varepsilon = 28$ K, and $q_C = 0.0$ [26]

For different species, the cross parameters are calculated by using the Lorentz-Bertholet combination rules [27].

Table 1: Parameters of the molecular models and interactions.

| Adsorbate | $\varepsilon$ (Kcal/mol) | $\sigma$ (nm) | $q$ (e) |
|---|---|---|---|
| C in $CO_2$ | 0.054 | 0.280 | +0.70 |
| O in $CO_2$ | 0.157 | 0.305 | -0.35 |
| O in $H_2O$ | 0.155 | 0.317 | -0.8476 |
| H in $H_2O$ | 0.0 | n/a | +0.4238 |

The GF substrate is created as follows. We first assemble four identical bilayer graphene flakes with 144 carbon atoms each in a cubic box of size 4x4x1 nm. Keeping each flake as a rigid body, they are "shaken" by running an MD simulation at T = 300 K. Then, we made 32 identical copies and positioned them on a squared lattice of 4x4x2 sites. The substrate obtained in this way has 18,432 carbon atoms and size of 16x16x2 nm (See Fig.2). The substrate has channels of width between 0.5 and 0.8 nm and a microporosity of 68% [5]. We have shown by Monte Carlo simulations that this substrate aligns with experimental features of activated carbons. [5].

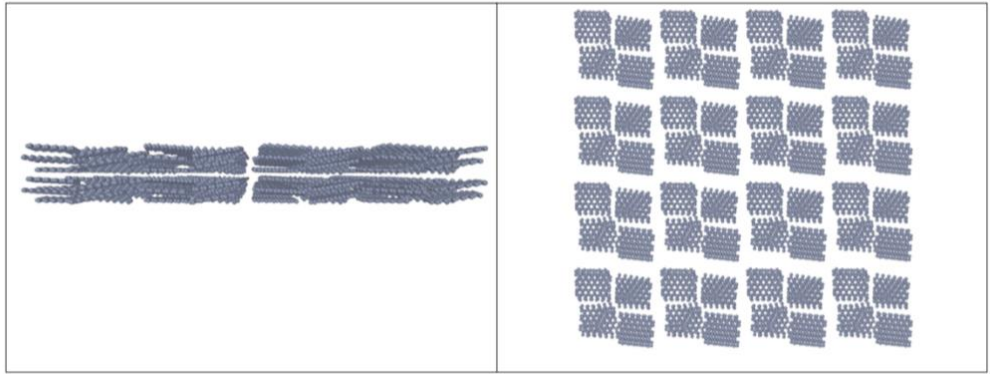

Figure 2: Graphene-flakes substrate: front view (left panel), and top view (right panel).

## 3. Results

Our simulations differ substantially from adsorption experiments. In the simulation cell, the pressure is not held constant; instead, the number of molecules is fixed. As molecules become adsorbed, the equilibrium vapor pressure therefore decreases. Each simulation runs for 10 million steps (approximately 5 nanoseconds), and equilibrium is not necessarily achieved within this timeframe.

Rather than computing the number of molecules adsorbed—such as by counting molecules within a region near the substrate—we evaluate the adsorption energy, $E_a$. This energy represents the interaction energy per particle between the molecules and the substrate. A more negative value of $E_a$ indicates stronger adsorption, while a less negative value corresponds to weaker adsorption. Thus, $E_a$ provides a measure of adsorption that does not depend on an arbitrary choice of the region defined as the adsorbed layer.

We define the reduced energy, $E'$ by scaling the magnitude of the adsorption energy with the LJ energy parameter between the molecule $i$ and a carbon atom, $\varepsilon_{iC}$,

$$E' = |E_a|/\varepsilon_{iC}$$

This reduced energy can be interpreted as an effective number of carbon atoms interacting with each adsorbed molecule.

Our main results for *E'* are presented in Fig. 3, which shows the reduced energy of $CO_2$ as a function of simulation time for temperatures ranging from 250 K to 375 K. For each temperature, we include the results of two simulations. The blue curves correspond to adsorption from vapors containing equal parts $CO_2$ and water (wet vapors), while the red curves correspond to adsorption from pure $CO_2$ vapor (dry vapors). We observe that the final reduced energy is higher for wet vapors at all temperatures except 375 K. As expected, adsorption decreases with increasing temperature.

We also compare the initial rate of adsorption, calculated as the slope of the linear trend lines over the first nanosecond. We find that the adsorption rate is higher for wet vapors than for dry vapors in all cases except at T = 300 K. This suggests that the presence of water may enhance $CO_2$ adsorption, with the effect being particularly pronounced at 350 K.

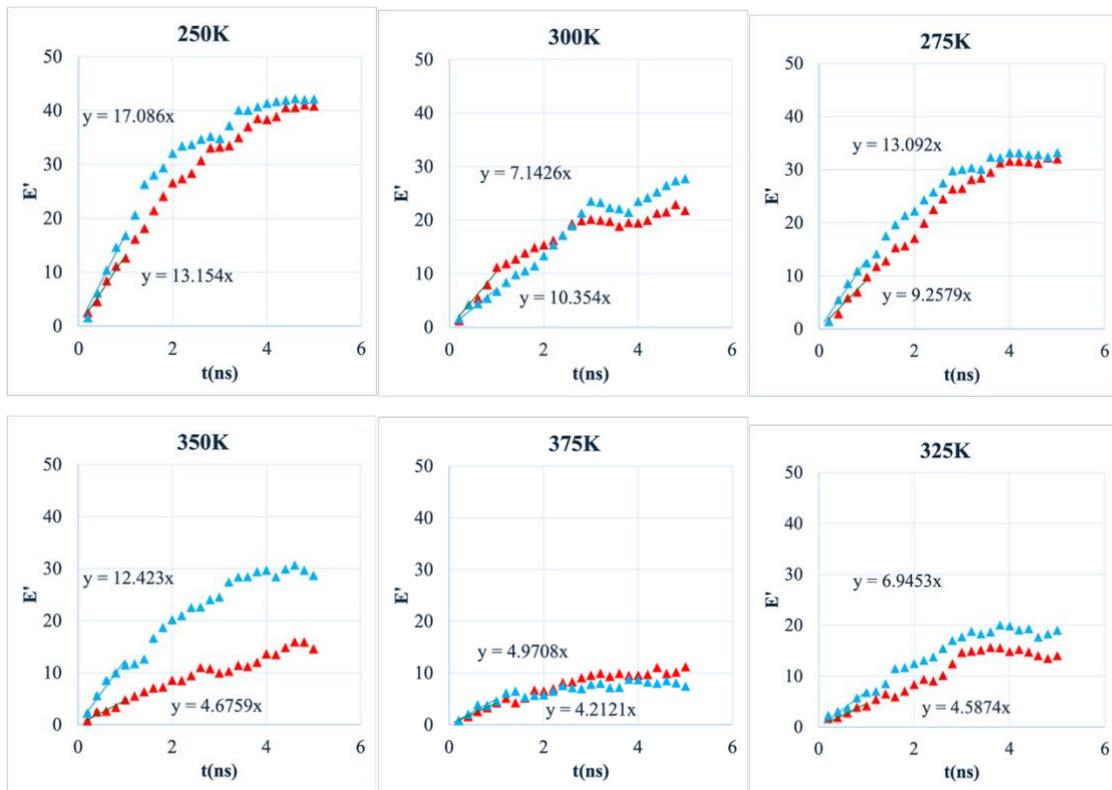

Figure 3: Reduced energy E' of $CO_2$ for temperatures from 250K to 375K, for wet (blue) and dry (red) vapors. The solid lines show linear trend-lines for the first-nanosecond data; the equations are included.

For all temperatures studied, CO2 is adsorbed both faster and more strongly than water, as shown in Fig. 4, which compares the reduced energy of the two molecules across all temperatures.

Although all adsorption sites are accessible to both species, we observe that CO2 predominantly occupies the central region of the substrate, whereas water is also adsorbed along the outer regions. This behavior is evident in Fig. 5, which shows the fraction adsorbed as a function of the z-coordinate. The substrate is located at the center of the simulation cell and spans a thickness of $\Delta z$ =2nm (see Fig. 1)

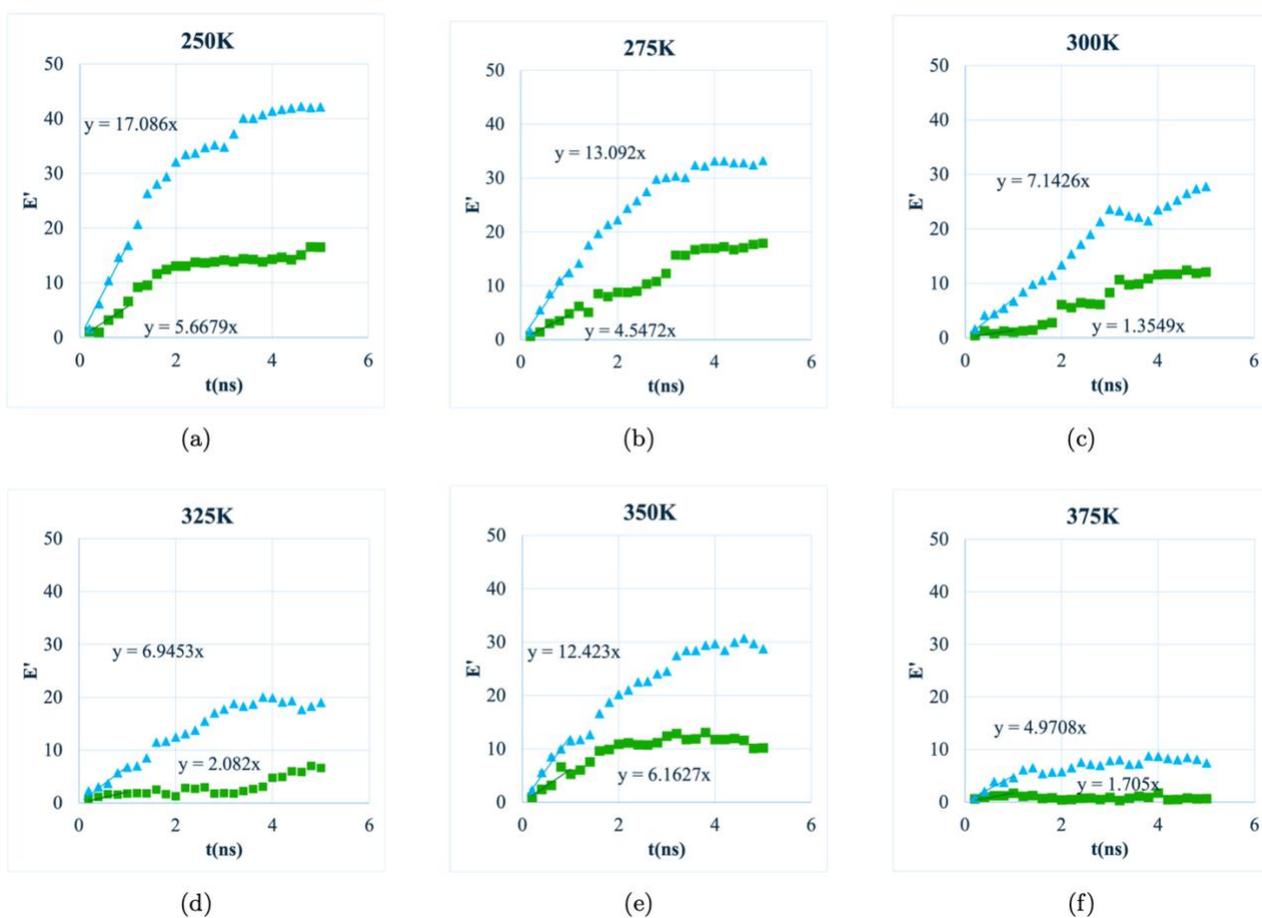

Figure 4: Reduced energy E' of CO2 (blue triangles) and water (green squares) for temperatures from 250K to 375K. The solid lines show linear trend-lines for the first-nanosecond data; the equations are included.

At the temperatures considered, water molecules rapidly form clusters. We quantify this clustering by computing the water–water interaction energy, $E_{WW}$, which reflects the extent of cluster formation. Fig. 6 shows $E_{WW}$ as a function of simulation time for several temperatures, revealing a high rate of clusterization except at the highest temperatures. Cluster formation and adsorption occur simultaneously.

As water clusters form, they are attracted to the substrate as "supermolecules." In doing so, they interact with $CO_2$ molecules that are also being adsorbed. These water clusters appear to act as a kind of "water substrate," enhancing $CO_2$ adsorption. To test this idea, we calculated the difference in reduced $CO_2$ energy between wet and dry vapors,

$$\Delta E'_{CO_2} = E'_{wet} - E'_{dry}$$

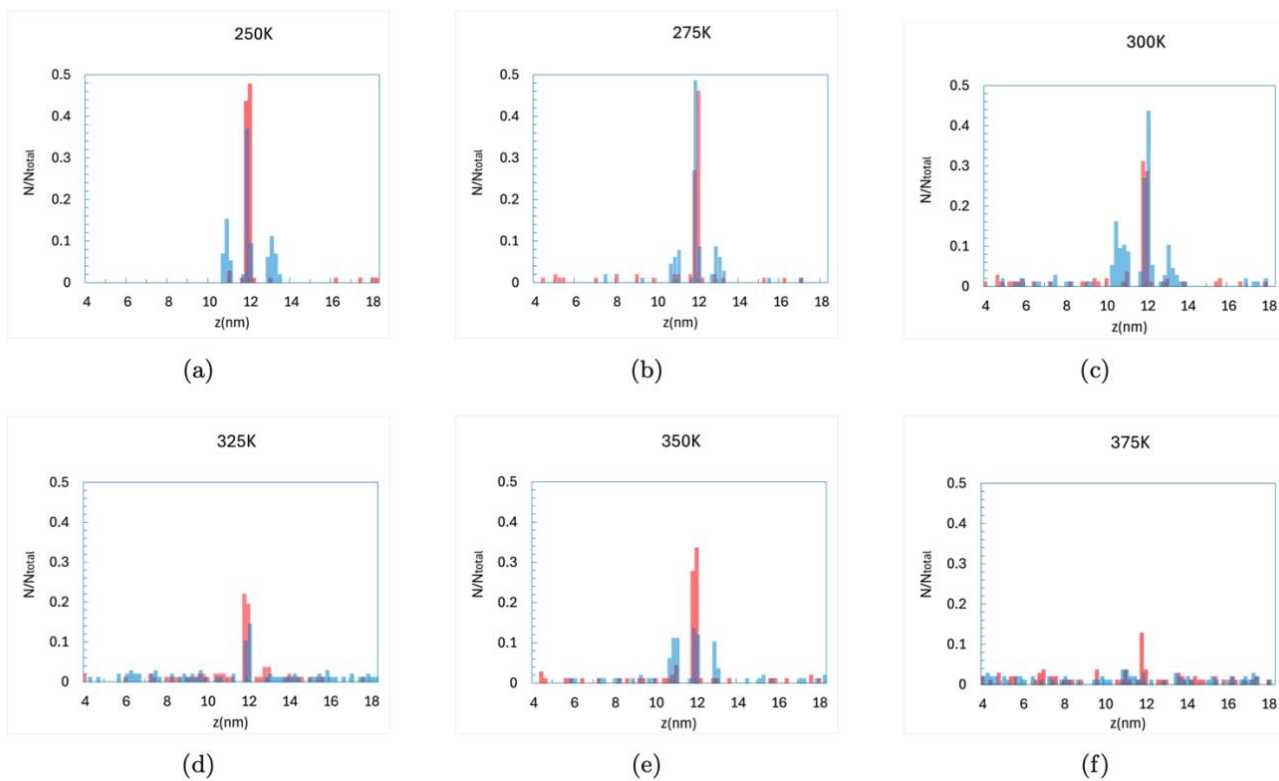

Figure 5: Fraction of $CO_2$ (red) and water (blue) molecules as a function of the coordinate z.

In Fig. 7 we compare $\Delta E'_{CO_2}$ with the $CO_2$-water interaction energy and the water–water interaction energy for temperatures of 250 K, 300 K, and 350 K. We observe that the largest $\Delta E'_{CO_2}$ occurs at 350 K, where both water clusterization and $CO_2$–water interactions are strongest.

Finally, we examine the effect of the substrate on cluster formation. Fig. 8 shows the water–water interaction energy with and without the substrate present. Our results indicate that the substrate inhibits the formation of water clusters.

## 4. Summary and Conclusions

In this work, we investigated the adsorption of mixed $CO_2$–water vapors on a graphene-flakes substrate using molecular dynamics simulations. By analyzing adsorption through the reduced energy, we obtained a robust, geometry-independent measure of adsorption strength that allowed comparison across species, temperatures, and vapor compositions.

Our results show that $CO_2$ adsorbs more strongly and more rapidly than water over the entire temperature range studied. For all temperatures below 375 K, adsorption from wet vapors leads to more negative (stronger) $CO_2$ reduced energies than adsorption from dry vapors. This indicates that the presence of water enhances $CO_2$ uptake under these conditions.

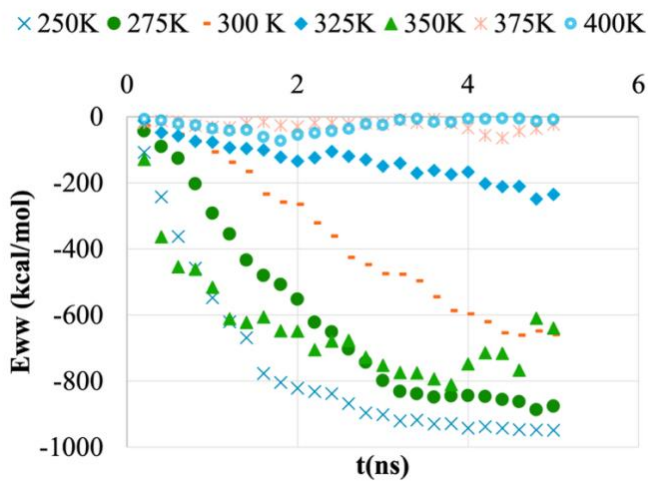

Figure 6: Energy of interaction between water molecules at temperatures from 250K to 400K.

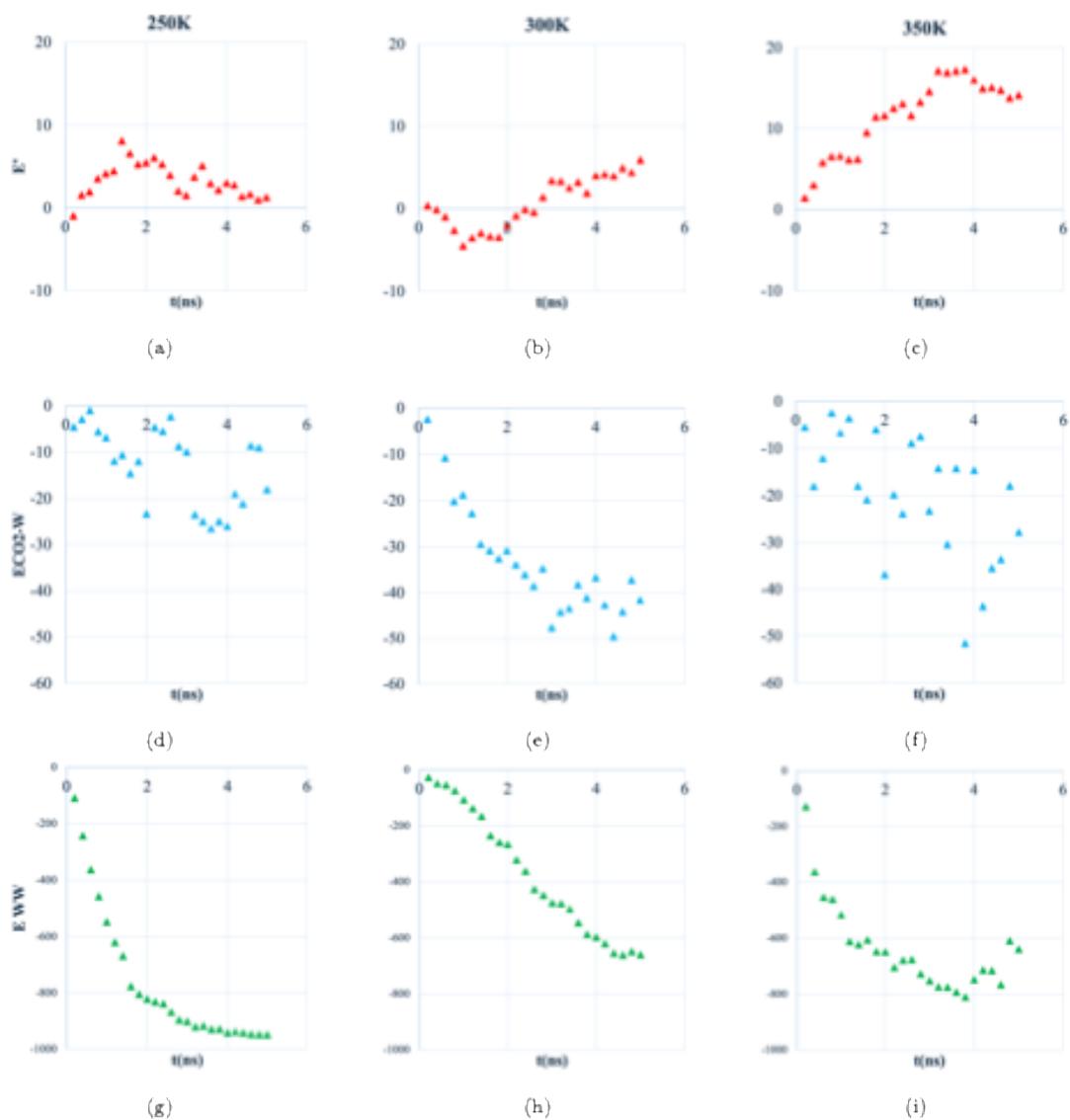

Figure 7: Top row: CO2-reduced energy difference between wet vapors and dry vapors; Center row: Energy of interaction between CO2 and water molecules (in kcal/mol). Bottom row: Water-water interaction energy (in kcal/mol).

The simulations suggest a mechanistic explanation for this enhancement. At moderate temperatures (250–350 K), water molecules form clusters that act as "supermolecules", which are themselves attracted to the substrate. As these clusters approach the GF surface, they interact with $CO_2$ molecules and effectively facilitate $CO_2$ adsorption. The correlation between the CO2–water interaction energy, water–water interaction energy, and the difference in reduced CO2 energy between wet and dry vapors shows that this cooperative effect is strongest at 350 K, where water clusterization and CO2–water interactions are maximized.

We also found that the GF substrate inhibits water cluster formation, as water–water interaction energies are significantly lower in the presence of the substrate than in its absence. This suggests that the porous graphene-flakes matrix disrupts large water aggregates while still allowing cluster-mediated enhancement of CO2 adsorption.

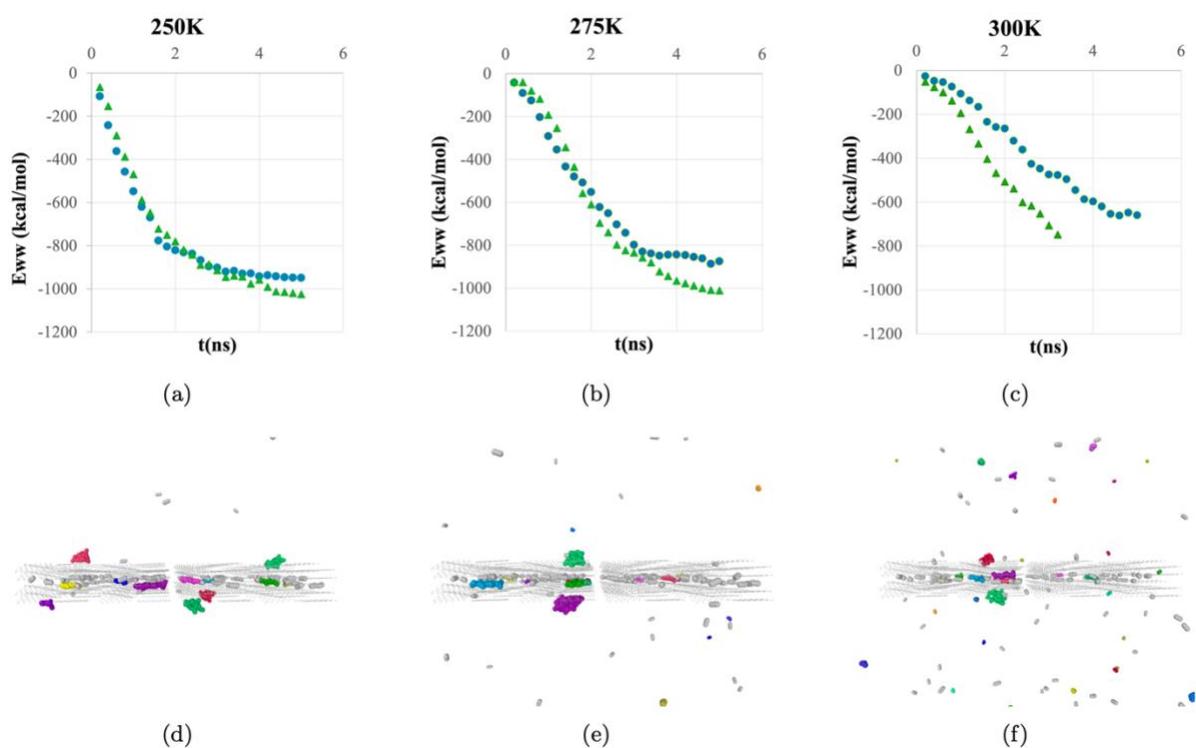

Figure 8: Top row: Energy of interaction between water molecules with (circles) or without (triangles) the substrate, at temperatures 250K, 275K and 300K. In the lower row, screenshots showing the clusters at t=5ns at each temperature.

Overall, our results indicate that graphene-flakes substrates not only preferentially adsorb CO2 over water but may also benefit from humidity at intermediate temperatures, where limited water clustering enhances CO2 uptake. These findings suggest that GF-based materials, inspired by activated carbons, could perform effectively under realistic, humid conditions and may offer a viable path toward efficient carbon capture from mixed vapors.

**Declaration of competing interest**

The authors declare that they have no known competing financial interests or personal relationships that could have appeared to influence the work reported in this paper.

**Acknowledgements**

This work used SDSC Expanse GPU at San Diego Supercomputer Center through allocation PHY230043 from the Advanced Cyberinfrastructure Coordination Ecosystem: Services Support (ACCESS) program, which is supported by U.S. National Science Foundation grants 2138259, 2138286, 2138307, 2137603, and 2138296.